\documentclass{desyproc}
\renewcommand{\v}[1]{\boldsymbol{#1}}		

\begin{document}
\title{Searching for Axion Dark Matter in Atoms:\\~Oscillating Electric Dipole Moments and Spin-Precession Effects}

\author{{\slshape  Benjamin M.~Roberts$^{1}$, Yevgeny V.~Stadnik$^1$, Victor V.~Flambaum$^{1,2}$, Vladimir A.~Dzuba$^1$}\\[1ex]
$^1$~School of Physics, University of New South Wales, Sydney 2052, Australia\\
$^2$~Mainz Institute for Theoretical Physics, Johannes Gutenberg University Mainz, D 55122 Mainz, Germany}

\contribID{Roberts\_Benjamin\_Axions}

\confID{11832}  
\desyproc{DESY-PROC-2015-02}
\acronym{Patras 2015} 
\doi  

\maketitle

\begin{abstract}
We propose to search for axion dark matter via the oscillating electric dipole moments that axions induce in atoms and molecules. These moments are produced through the intrinsic oscillating electric dipole moments of nucleons and through the $P,T$-violating nucleon-nucleon interaction mediated by pion exchange, both of which arise due to the axion-gluon coupling, and also directly through the axion-electron interaction. Axion dark matter may also be sought for through the spin-precession effects that axions produce by directly coupling to fermion spins.

\end{abstract}

\section{Introduction}
Astrophysical observations indicate that the matter content of the Universe is overwhelmingly dominated by dark matter (DM), the energy density of which exceeds that of ordinary matter by a factor of five. In order to explain the observed abundance of DM, it is reasonable to expect that DM interacts non-gravitationally with ordinary matter. Searches for weakly interacting massive particle (WIMP) DM, which look for the scattering of WIMPs off nuclei, have not yet produced a strong positive result. Further progress with these traditional searches is hindered by the observation that the sought effects are fourth-power in the underlying interaction strength between DM and Standard Model (SM) matter, which is known to be extremely small. 

We propose to search for other well-motivated DM candidates that include the axion, which may also resolve the strong \emph{CP} problem of Quantum Chromodynamics (QCD), by exploiting effects that are first-power in the interaction strength between the axion and SM matter (by contrast, haloscope \cite{ADMX2010} and helioscope \cite{IAXO2014} methods look for second-power effects, while light-shining-through-wall methods \cite{Bahre2013} look for fourth-power effects). We focus on the oscillating electric dipole moments (EDMs) and spin-precession effects that axions induce in atoms and molecules. There is strong motivation to search for axions in atomic and related systems via such signatures --- to date, static EDM measurements in atoms, molecules and ultracold neutrons have served as sensitive probes of new physics beyond the Standard Model (see e.g.~the reviews \cite{Ginges2004Review,Pospelov2005ReviewEDM,Roberts2015Review}), while searches for sidereal spin-precession effects with atoms and ultracold neutrons have placed stringent limits on \emph{CPT}- and Lorentz-invariance-violating models (see e.g.~Ref.~\cite{Kostelecky2011} for an overview).


\section{Axion dark matter}
Axions produced by the vacuum misalignment mechanism are cold with no pressure. Furthermore, if axions are sufficiently light and weakly interacting, then they may survive until the present day and reside in the observed galactic DM haloes (where they have become virialised over time with $v_{\textrm{virial}} \sim 10^{-3}$). The number density of ultralight (sub-eV mass) axions per de Broglie volume readily exceeds unity, $n_a / \lambda_{\textrm{dB}}^3 \gg 1$, meaning that axions behave as a coherently oscillating classical field, $a(t) \simeq a_0 \cos(m_a t - \v{p}_a \cdot \v{r})$ on time scales less than $\tau_{\textrm{coh}} \sim 2\pi / m_a v_{\textrm{virial}}^2$ and on length scales less than $l_{\textrm{coh}} \sim 2\pi/m_a v_{\textrm{virial}}$. Thus the couplings of an oscillating galactic axion field to SM particles produce a number of oscillating signatures which can be sought for experimentally. As we will see below, the particularly interesting signatures are those where the observables scale as $\mathcal{O} \propto 1/f_a$ with the axion decay constant $f_a$. 

The axion couplings to SM particles that are of most interest are the following:
\begin{align}
\label{Axion_couplings}
\mathcal{L}_{\textrm{int}} = \frac{a}{f_a} \frac{g^2}{32\pi^2} G\tilde{G}  ~ - \sum_{f=e,n,p} \frac{C_f}{2f_a} \partial_\mu a ~ \bar{f} \gamma^\mu \gamma^5 f ,
\end{align}
where the first term represents the coupling of the axion field to the gluonic field tensor $G$ and its dual $\tilde{G}$, 
and the second term represents the coupling of the derivative of the axion field to the fermion axial-vector currents. $C_f$ are dimensionless model-dependent coefficients. Typically, $|C_n| \sim |C_p| \sim 1$ in models of the QCD axion \cite{Srednicki1985axion}. Within the DFSZ model, where the tree level coupling of the axion to the electron is non-vanishing, $|C_e| \sim 1$ \cite{Srednicki1985axion}. However, within the KSVZ model, $|C_e| \sim 10^{-3}$, since the tree level coupling vanishes and the dominant effect arises at the 1-loop level \cite{Srednicki1985axion}. For more generic axion-like pseudoscalar particles, the coefficients $C_f$ are essentially free parameters, and the coupling to gluons is generally presumed absent.

\:
\:
\:

\begin{figure}[h!]
\begin{center}
\includegraphics[width=4.4cm]{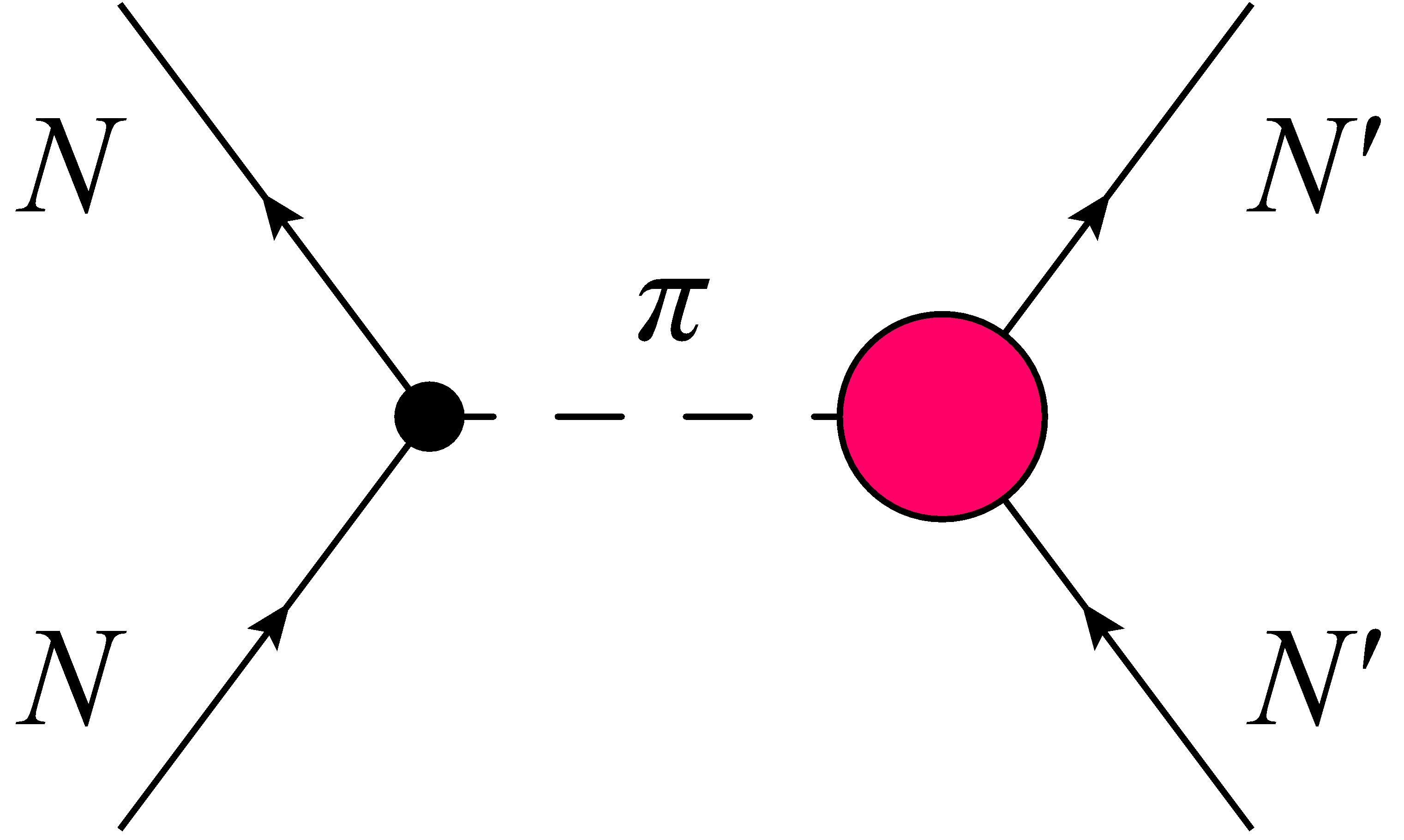}
\caption{Main process responsible for the induction of oscillating $P$,$T$-odd nuclear electromagnetic moments by an oscillating axion field. The black vertex on the left is due to the usual strong $P$,$T$-conserving $\pi N N$ coupling ($g_{\pi N N} = 13.5$), while the magenta vertex on the right is due to the axion-induced $P$,$T$-violating $\pi N N$ coupling ($\bar{g}_{\pi N N} \approx 0.027 \cos(m_a t) ~ a_0/f_a $) \cite{Stadnik2014axions}.} 
\label{fig:Pion_nucleon_PT-odd}
\end{center}
\end{figure}

\textbf{Oscillating P,T-violating nuclear electromagnetic moments.} ---
The coupling of an oscillating axion field to the gluon fields, which is described by the first term in Eq.~(\ref{Axion_couplings}), induces an oscillating EDM of the neutron \cite{Graham2011,Stadnik2014axions}:
\begin{equation}
\label{NEDM_axion}
d_n(t) \approx 1.2 \times 10^{-16} ~\frac{a_0}{f_a} \cos(m_a t) ~ e \cdot \textrm{cm} ,
\end{equation}
which in turn induces oscillating \emph{P},\emph{T}-violating nuclear electromagnetic moments. In nuclei, a second and more efficient mechanism exists for the induction of oscillating electromagnetic moments by axions --- namely, the \emph{P},\emph{T}-violating nucleon-nucleon interaction that is mediated by pion exchange, with the axion field supplying the oscillating source of \emph{P} and \emph{T} violation at one of the $\pi N N$ vertices \cite{Stadnik2014axions} (Fig.~\ref{fig:Pion_nucleon_PT-odd}).

\:
\:
\:

\textbf{Oscillating atomic and molecular electric dipole moments.} ---
Axion-induced oscillating $P$,$T$-odd nuclear electromagnetic moments can in turn induce oscillating EDMs in atoms and molecules. In diamagnetic species ($J=0$), only oscillating nuclear Schiff moments (which require $I \ge 1/2$) produce an oscillating atomic/molecular EDM (oscillating nuclear EDMs are effectively screened for typical axion masses, as a consequence of Schiff's theorem \cite{Schiff1963}). Two atoms that are of particular experimental interest are $^{199}$Hg and $^{225}$Ra, for which the axion induces the following oscillating EDMs \cite{Stadnik2014axions}:
\begin{align}
\label{AtomicEDM_axion-Hg}
d(^{199}\textrm{Hg}) \approx -1.8 \times 10^{-19} ~\frac{a_0}{f_a} \cos(m_a t) ~ e \cdot \textrm{cm} , \\
d(^{225}\textrm{Ra}) \approx 9.3 \times 10^{-17} ~\frac{a_0}{f_a} \cos(m_a t) ~ e \cdot \textrm{cm} ,
\end{align}
with the large enhancement in $^{225}$Ra compared with $^{199}$Hg due to both collective effects and small energy separation between members of the parity doublet concerned, which occurs in nuclei with octupolar deformation and results in a significant enhancement of the nuclear Schiff moment \cite{Auerbach1996,Auerbach1997}. A possible platform to search for the oscillating EDMs of diamagnetic atoms in ferroelectric solid-state media has been proposed in Ref.~\cite{CASPER2014}.

Paramagnetic species ($J \ge 1/2$) offer more rich possibilities. Firstly, axion-induced oscillating nuclear magnetic quadrupole moments (which require $I \ge 1$) also produce an oscillating atomic/molecular EDM \cite{Roberts2014long}, which is typically larger than that due to an oscillating nuclear Schiff moment (since magnetic quadrupole moments are not subject to screening of the applied electric field by atomic/molecular electrons). Secondly, an entirely different mechanism exists for the induction of oscillating EDMs in paramagnetic species, through the direct interaction of the axion field with atomic/molecular electrons via the second term in Eq.~(\ref{Axion_couplings}). The $\mu = 0$ component of this second term mixes atomic/molecular states of opposite parity (with both imaginary and real coefficients of admixture), generating the following oscillating atomic EDM (due to the real coefficients of admixture) in the non-relativistic approximation for an $S_{1/2}$ state \cite{Stadnik2014axions}:
\begin{equation}
\label{atomic_EDM_para}
d_a (t) \sim - \frac{C_e a_0 m_a^2 \alpha_s}{f_a \alpha} e \cos(m_a t) ,
\end{equation}
where $\alpha_s$ is the static scalar polarisability. Fully relativistic Hartree-Fock atomic calculations are in excellent agreement with the scaling $d_a \propto \alpha_s$ in Eq.~(\ref{atomic_EDM_para}) \cite{Roberts2014long,Roberts2014prl}. The imaginary coefficients of admixture in the perturbed atomic wavefunction produce \emph{P}-violating, \emph{T}-conserving effects in atoms, while the analogous imaginary coefficients of admixture in the perturbed nuclear wavefunction (due to the axion-nucleon interaction via the $\mu = 0$ component of the second term in Eq.~(\ref{Axion_couplings})) produce \emph{P}-violating, \emph{T}-conserving nuclear anapole moments \cite{Stadnik2014axions,Roberts2014long,Roberts2014prl}.

\:
\:
\:

\textbf{Oscillating spin-precession effects.} ---
The coupling of an oscillating axion field to the fermion axial-vector currents produce the following time-dependent non-relativistic potential for a spin-polarised source, via the $\mu = 1,2,3$ components of the second term in Eq.~(\ref{Axion_couplings}):
\begin{equation}
\label{potential_axion-wind}
H_{\textrm{int}} (t) = \sum_{f=e,n,p} \frac{C_f a_0}{2 f_a} \sin(m_a t) ~ \v{\sigma}_f \cdot \v{p}_a ,
\end{equation}
which give rise to spin-precession effects \cite{Stadnik2014axions,Flambaum2013Patras,Graham2013}. Deformation of the axion field by the gravitational field of a massive body produces also a time-dependent potential of the form $H'_{\textrm{int}} (t) \propto C_f a_0 \sin(m_a t) / f_a ~ \v{\sigma}_f \cdot \v{\hat{r}}$, which is directed towards the centre of the gravitating body \cite{Stadnik2014axions}. These spin-precession effects can be sought for using a wide range of spin-polarised systems, including atomic co-magnetometers, ultracold neutrons and torsion pendula. The nucleon spin contents for nuclei of experimental interest have been performed in Ref.~\cite{Stadnik2015NMBE} for the accurate interpretation of laboratory measurements.

\section*{Acknowledgments}
This work was supported by the Australian Research Council. B.~M.~R.~and V.~V.~F.~are grateful to the Mainz Institute for Theoretical Physics (MITP) for its hospitality
and support.


\begin{footnotesize}

\end{footnotesize}


\end{document}